\documentclass[useAMS,usenatbib]{mn2e_mod}

\usepackage{comment}

\usepackage[latin1]{inputenc}
\usepackage{latexsym,textcomp}
\usepackage{fancyhdr}
\usepackage{lscape}
\usepackage{verbatim}
\usepackage{subfigure}
\usepackage{multicol}
\usepackage{blindtext}
\usepackage[]{graphicx}

\usepackage[pdftitle={Modal noise prediction in fibre-spectroscopy},
            pdfsubject={Modal noise}
             pdfauthor={Ulrike Lemke}]{hyperref}
 \usepackage[authoryear]{natbib}

\newcommand{\vs}[1] {\vspace{#1}}

\newcommand{\ns} \normalsize

\newsavebox{\fmbox}
\newenvironment{fmpage}[1]
{\begin{lrbox}{\fmbox}\begin{minipage}{#1}}
{\end{minipage}\end{lrbox}\fbox{\usebox{\fmbox}}} 

\newcounter{satzcounter} 

\newcounter{defcounter} 

\title{Modal noise prediction in fibre-spectroscopy I: Visibility and the coherent model}

\author[U. Lemke et al.]{U.~Lemke$^1$\thanks{E-mail: ulrike.lemke@durham.ac.uk}, J. Corbett$^1$, J. Allington-Smith$^1$, and G. Murray$^1$ \\
  $^1$ Durham University, South Rd, Durham DH1 3LE}

\begin{document}

  \maketitle
  \begin{abstract}
Fibre modal noise occurs in high spectral resolution, high signal-to-noise applications. It imposes fundamental limits on the photometric accuracy of state-of-the-art fibre-spectrograph systems. In order to maximize the performance of current and future instruments it is therefore essential to predict fibre modal noise. To attain a predictive model we are using a dual approach, bringing theoretical assumptions in line with the experimental data  obtained using a test-bench spectrograph. We show that the task of noise prediction can be reduced to determining the \textit{visibility} of the modal pattern which can be measured at the detector plane. Subsequently, the visibility-dependence of essential parameters is presented. This work will soon provide a basis for prediction of modal noise limitations in fibre-coupled spectrograph designs.
\end{abstract}  

\section{Introduction}
\subsection{Modal noise}
Modal noise is a measurement uncertainty caused by inherent statistical properties of an optical fibre. It can be ascribed to an interference effect at the fibre output, occurring when light of limited bandwidth is detected. Hence, modal noise is particularly important for high-resolution applications where central wavelength determination, spectral line ratios and shapes play a crucial role. This paper concentrates on the effect of modal noise on the photometric accuracy in a measurement. Generally speaking, any mode filtering process affects the throughput of the combined fibre-spectrograph system. In particular, beam truncation (compare fig. \ref{fig:speckle_aperture}) will limit the achievable signal-to-noise of the instrument.\\
As telescope apertures increase in size and ever higher spectral resolving powers become available, higher signal-to-noise ratios (SNR) can be achieved and the light at the detector becomes more coherent. The employment of optical fibres plays an important role here in order to achieve the required instrument stability and spectral resolving power (see \cite{Heacox_1988} and \cite{Allington_Smith_2006}). Unlike in conventional slit spectroscopy, fibres only partially retain spatial and angular information. For differing seeing conditions the light that is coupled into the fibre largely causes the same spectrograph response and thus stabilizes the signal; A property that is referred to as \textit{scrambling} in the literature (see \cite{Avila_2008}). With these advantages and the increasing requirements on instrument precision, the modal noise issue will continue to gain in prominence. It is therefore essential to predict modal noise in order to determine the performance of the complete instrument, allowing for optimization and precautionary measures that help reducing modal noise before it becomes a fundamental constraint to the instrument.\\
\cite{Epworth_1979} first noted that the modal properties of fibres will limited the accuracy of intensity measurement. Inherent fibre properties cause a speckle pattern at the fibre exit which is subject to changes in the fibre condition, e.g. stress or temperature shifts. In a subsequent paper by \cite{Rawson_Goodman_Norton_1980} the bandwidth-dependence of the speckle contrast was studied: A lower coherence (e.g. in low-resolution spectrographs) results in diminished speckle contrast (low visibility). This theory predicts a strong length-dependence which was disproven for fibre lengths $<$\,$10^2$\,m by \cite{Baudrand_Walker_2001, Corbett_2006}.
Although coherence effects are a key parameter, attention has to be paid to the statistical properties of the speckles and their effect on modal noise. As was recently pointed out by \cite{Grupp_2003} spatially filtering apertures are often deployed, e.g. fibre-slit at spectrograph entrance or overfilled grating, initially aimed at high spectral resolution at low grating cost - but with disregard to limiting modal noise.\\
The first investigation to address aperture clipping was conducted by \cite{Hill_1980}, and soon followed by \cite{Goodman_Rawson_1981}.
This work showed that modal noise is caused by mode filtering processes but is restricted to the extreme case of highly coherent light. The first work to combine coherence effects and speckle statistics into a concise model is reported by \cite{Kanada_1983}. Their experimental results show good agreement with theoretical predictions, but the investigation concentrates on high connection losses, i.e. strong beam vignetting. Moreover, the considered visibility regime of $>$\,10\% is significantly higher than relevant in astronomical spectroscopy (5\% and less in this work). The first prediction of the wavelength-dependent spectrograph response was achieved by  \cite{Chen_2006} for few-mode fibres. Their mode-propagation simulation is in excellent agreement with experimental findings, allowing for precise modal noise prediction. But the multi-mode case is  more complicated and already the absence of strong fibre-length dependence suggests that our theoretical picture of light propagation in optical fibres is not yet sufficiently developed.\\
Today's instruments are already meeting challenging requirements, among these the most outstanding are instruments dedicated for exoplanet research using the radial velocity method. High-resolution systems like HARPS and ESPRESSO are now striving to detect Earth-like planets, but the unprecedented precision required here is found to be restricted (amongst other effects) by modal noise (see \cite{Wilken_2010, Boisse_2010}). \\ 
Stellar observations that measure metallicities or make use of analytical tools like the Stark effect for temperature determination will be affected. Also studies on interstellar material measuring ionization levels will ultimately be limited by modal noise. Furthermore, for solar observations working at high resolution, high signal-to-noise levels are affected by modal noise (e.g. high-precision-polarimetry).

\subsection{Scope of this investigation}
This paper addresses modal noise in the parameter range of spectrographs dedicated to astronomical applications. We concentrate on intermediate to low aperture restrictions and visibilities $<$\,10\%. In contrast to the literature discussed above, we consider the effect of modal noise for the low coherence case (in contrast to the investigations at high coherence found in literature) and for the fibre spectrum rather than just the fibre image.\\
A full theoretical picture of the effects involved, requires understanding phenomena like intermodal coupling, modal filtering and spatial mode distribution. Therefore, we feel it necessary that models with the objective of modal noise prediction need to be grounded on experimental results first.\\
We start out in section \ref{sec:theoretical_predictions} with the existing theoretical background of modal noise. A first simplification is possible with the assertion (which will be experimentally verified), that modal noise can be described as the product of two independent parameters; a visibility and a coherent term. Hence, it can be separated into two sub-problems: The coherent speckle term is well understood from fibre and restricting aperture geometry, whereas visibility is mainly dependent on resolving power but also on the slit width. A simple model for visibility is derived from simplified ray propagation assumptions, outlining basic dependencies on spectrograph parameters.\\
For the experimental investigations a medium to high resolution ($R= \lambda/\Delta\lambda$ = 10\,000 - 200\,000) Echelle spectrograph test-bench was built. Its design and the data-analysis are presented in section \ref{sec:exp}. The experimental findings (section \ref{sec:speck_stats}) support the above hypothesis about the modal noise factorization, the coherent part is proven to follow the theoretical predictions of \cite{Goodman_Rawson_1981}. The problem of modal noise prediction thus reduces to the problem of determining the visibility (section \ref{sec:visibility}). This fact can already be used to directly extract modal noise information from raw data of calibration sources. To enable direct prediction of visibility and modal noise, we furthermore investigate the dependence of visibility on spectrograph parameters such as resolving power and input focal ratio. Although some agreement between theoretical predictions and experimental data was shown, additional investigation needs to follow to refine the visibility-model and thus enable modal noise prediction. This work will be subject to a later paper (Lemke et al., in preparation).

\section{Theoretical predictions}\label{sec:theoretical_predictions}

\begin{figure} 
  \centering
  \includegraphics[width=.5\textwidth]{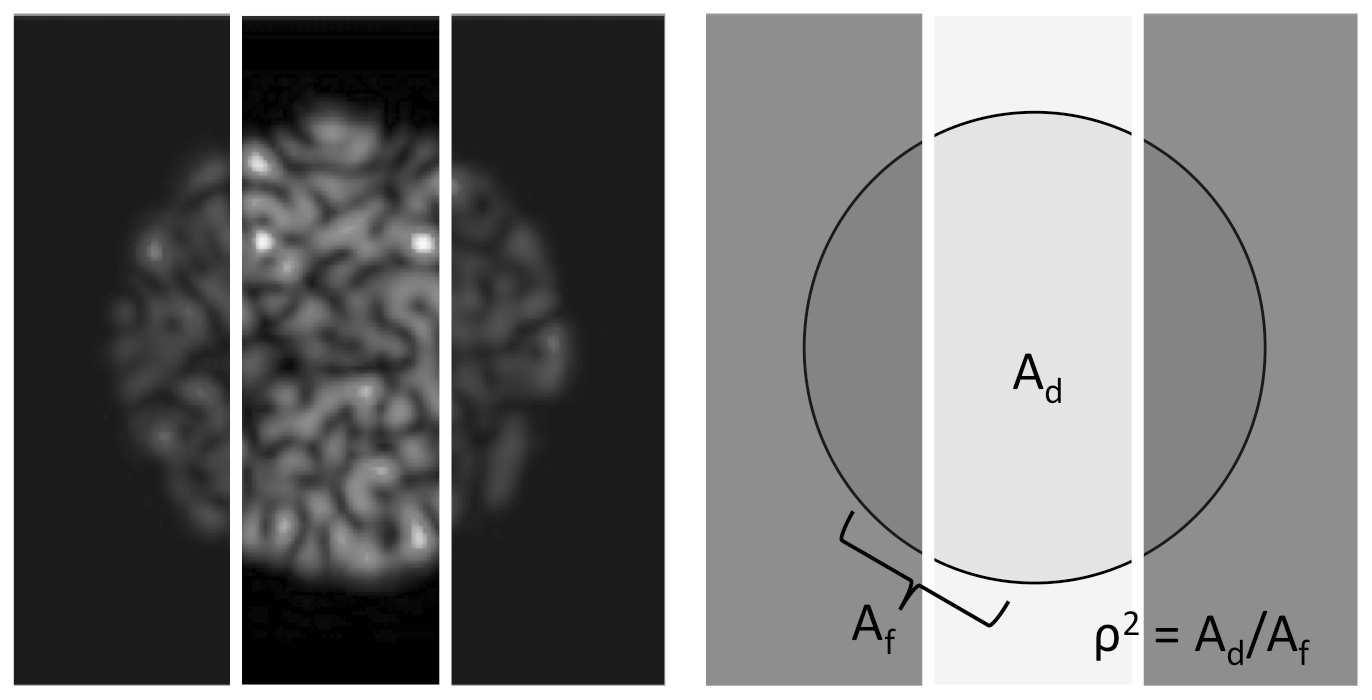}
    \caption{Modal noise originates from truncating the beam of an inhomogeneous, timely varying intensity distribution. The essential geometrical variable $\rho^2$ is defined by two parameters; $A_f$ is the area of the fibre image, whereas $A_d$ is the part of the illuminated area that is detected.}
  \label{fig:speckle_aperture}
\end{figure}	
In confined waveguides as represented here by the standard optical fibre, the electromagnetic field has to fulfill certain boundary conditions. As a result, only the fraction of the light that matches the potential walls can propagate through the fibre and is quantized into fibre modes. The typical fibre differs from the ideal circular symmetric waveguide - along the length of the fibre a variation of parameters like core-diameter and refractive index occurs. These statistical properties cause a change in symmetry and phase relation of the supported modes, a phenomenon that manifests in a speckle pattern as in fig. \ref{fig:speckle_aperture} visible in coherent light. Changing stress in the fibre (as for instance due to telescope tracking) causes the speckle pattern to change and in connection with mode filtering processes such as beam truncation causes modal noise.\\
The $S\!N\!R$ derives from theoretical considerations as follows: According to \cite{Goodman_Rawson_1981} the $S\!N\!R$ can be calculated from speckle statistics and geometrical considerations, given the number of speckles (which is equal to the number of excited modes $M$, see \cite{Daino_1979}) and the area fraction $\rho^2$ of the illuminated fibre end-face that is transmitted to the detector surface  (see fig. \ref{fig:speckle_aperture}):\\
 \begin{equation}\label{eq:GR}
S\!N\!R_{coh} =\rho\sqrt{\frac{M+1}{1-\rho^2}},
 \end{equation}
Note that the truncation described by $\rho^2$ is not necessarily generated by a slit at e.g. the spectrograph entrance, but also by overfilling the grating or any other spatially filtering process (e.g. inhomogeneous sensitivity of the CCD). The subscript ${coh} $ indicates that we consider the coherent case. Note that the question of how many modes are excited is not trivial. Although the mode volume $V$ for a given input illumination can be directly calculated (see e.g. \cite{Hill_1980}) from the fibre core radius $a$, wave number $k = 2\pi/\lambda$ and angle $\theta_{in}$ of the light cone feeding into the fibre 
 \begin{equation}\label{eq:mode_volume}
V = \frac{1}{2}\left(ak\theta_{in}\right)^2,
 \end{equation}
this is only an upper limit. $M$ very much depends on the fibre-launch geometry (axis-offset, angle), as was asserted by \cite{Corbett_2007}.\\
However, these considerations only hold true when considering coherent light. The speckle contrast is reduced when observing light of broad bandwidth and studying a large set of fibre modes, as is the case for fast coupled multimode fibres. Then the fibre is populated by a large number of modes, resulting in a broad range of propagation constants. This is taken into account by \cite{Kanada_1983}, dividing the pure speckle statistics expression by the visibility $v$, yielding
 \begin{equation}\label{eq:GR_KA}
S\!N\!R =\frac{1}{v}S\!N\!R_{coh}.
 \end{equation}

Here $v$ can be theoretically derived by taking into account the finite bandwidth of the optical source ($f_c$) and the phase delay between modes ($\sim\!f_s$):

\begin{equation}\label{eq:KA}
 v_{th} = \left[1+2(f_s/f_c)^2\right]^{-1/2}.
 \end{equation}
 
An approximation (see Appendix~A) helps understanding the structure of the equation and its dependence on spectrograph parameters, leading to the following result:\\
\begin{equation}\label{eq:KA2}
v_{th} = \left[1+2\left(\frac{L\theta^2}{2n_{c}R\lambda_c}\right)^2\right]^{-1/2},
\end{equation}
for fibre length $L$, numerical aperture $\theta$, core refractive-index $n_c$, spectrograph resolving power $R$ and  central wavelength $\lambda_c$.\\
We make no claim here that these simple assumptions lead to a model fully consistent with the experiment, but they should serve us as a preliminary theoretical framework. For instance it was shown by \cite{Baudrand_Walker_2001, Corbett_2006}, that the $S\!N\!R$ has no significant dependence on fibre length  for $L\leq$ 100\,m. This is possibly because the fibre does not terminate with a perfect quartz-air intersection. This so called \textit{end-effect} (\cite{Poppett}) may lead to a stronger mode mixing and thus less coherence between the modes, which would explain the diminished visibility. In the following $L$ will therefore be treated as a free parameter. Thus, $v$ remains a function of $\theta$, $n_c$, $R$, $\lambda$, we will concentrate here on $\theta$ and $R$ only.
In the limit of low visibilities it applies that $(L\theta^2)/(2n_{c}R\lambda_c)\gg1$, so that eq. (\ref{eq:KA2}) can be approximated by
\begin{equation}\label{eq:vis_est}
v_{th}\propto\theta^{-2}s^{-1},
\end{equation}
where $R$ is assumed to be proportional to the slit width $s$. We will see that this approximation is fulfilled for the $\theta$-case and fulfilled at least by trend for $s$, whereas the full dependence of visibility on the spectrograph resolving power $R$ will remain a matter for further investigations.\\

\section{Experimental setup}\label{sec:exp}

\subsection{Spectrograph setup}

The experimental apparatus consists of fibre input, fibre agitator, slit unit and spectrograph.\\

\begin{figure*}
    \begin{center}
	\subfigure[]{\label{fig:fiber_input}\includegraphics[width=0.5\textwidth]{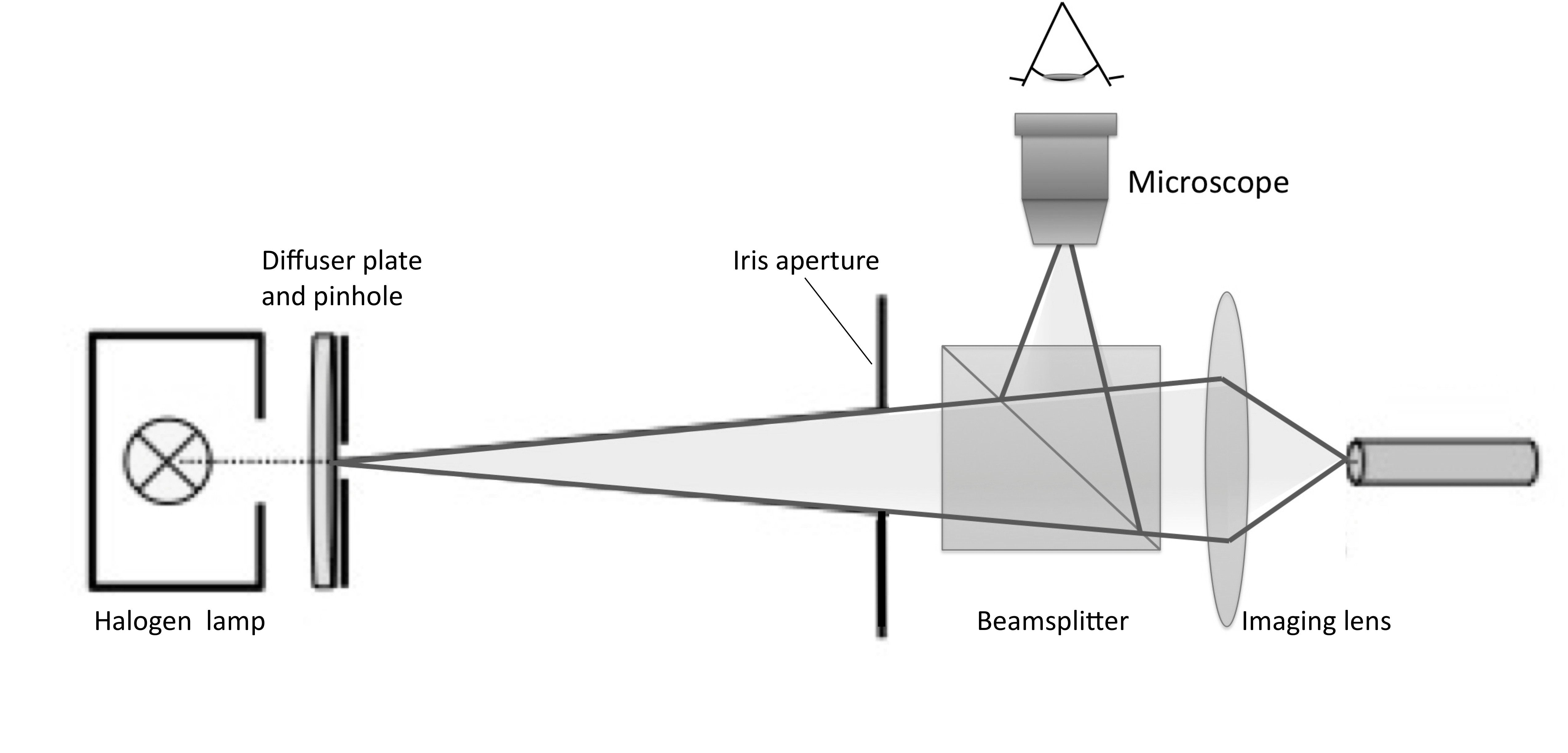}}
    \hfill
    	\subfigure[]{\label{fig:fibre_agitation}\includegraphics[width=0.4\textwidth]{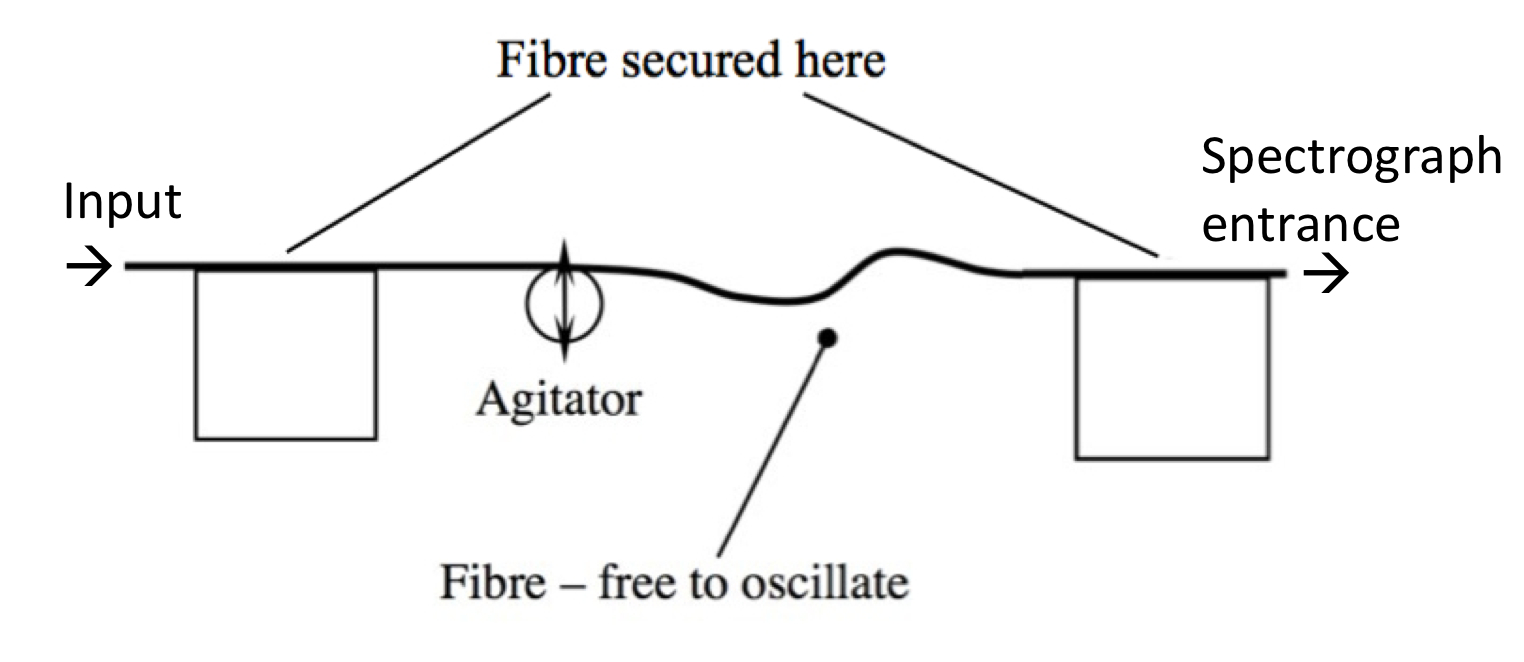}}
    \end{center}\caption{(a) Fibre input: The focus position can be controlled with the aid of a microscope in order to obtain optimal fibre illumination; (b) Fibre agitation scheme: The fibre is fixed at two points with a large free length in between. An agitator is used to produce large deflections in this region.}\label{fig:input_ag}
\end{figure*}

\begin{figure*} 
  \centering
  \includegraphics[width=.6\textwidth]{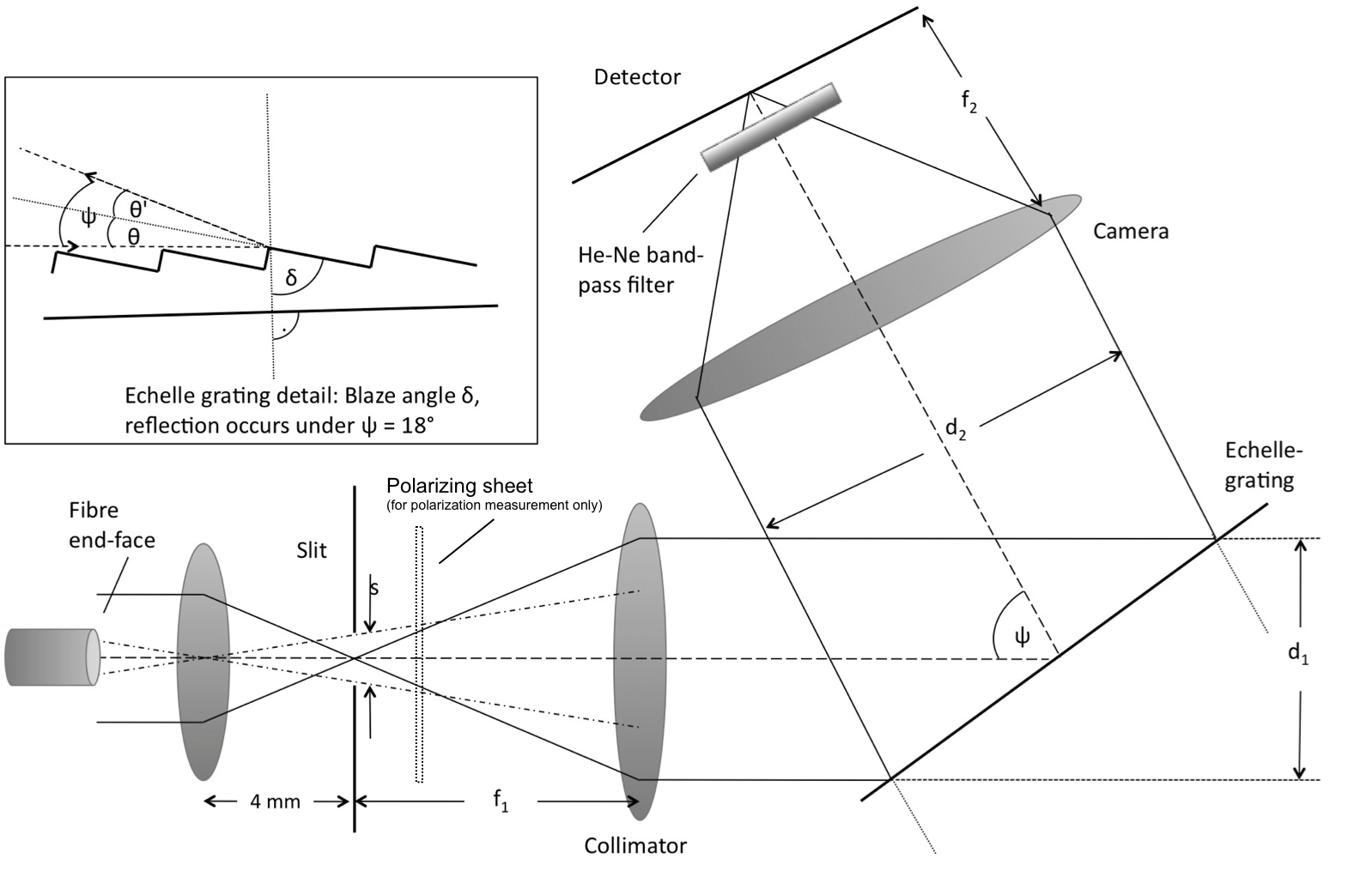}
    \caption{Schematic depiction of the Echelle spectrograph used for experimental investigations. The fibre far-field is imaged onto the slit plane which coincides with the spectrograph entrance. High spectral resolution is achieved by using the Echelle grating in Blaze configuration. Light is filtered through a narrow bandpass filter to avoid spectra overlapping at the CCD. (Slit and 4\,mm lens are not to scale, grating angle exaggerated.)}
  \label{fig:spec}
\end{figure*}

\begin{figure*} 
    \begin{center}
    \subfigure[]{\label{fig:diff_agschemes}\includegraphics[height = .20\textwidth, width=.5\textwidth]{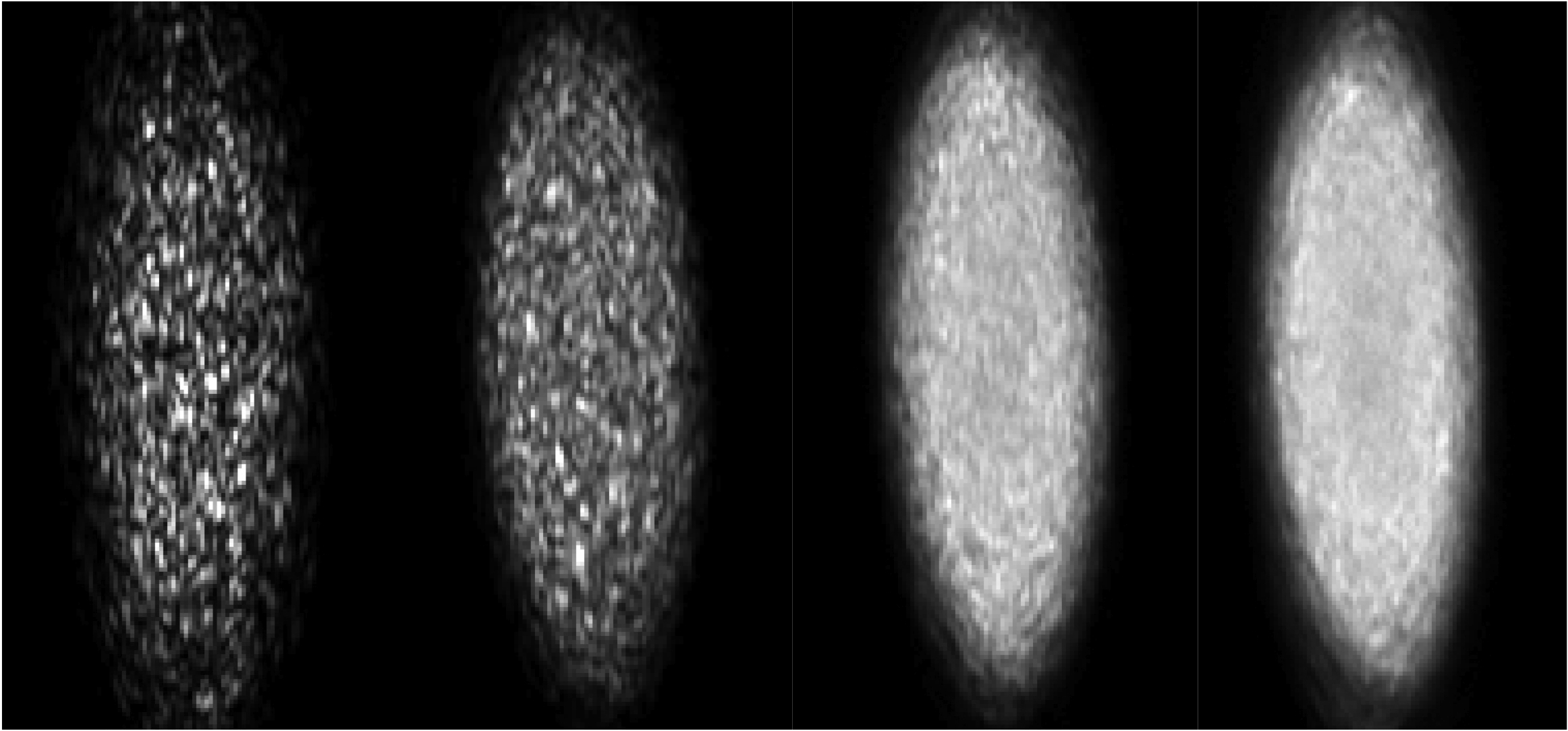}}
    \subfigure[]{\label{fig:other_rhosq} \includegraphics[height = .20\textwidth, width=.375\textwidth]{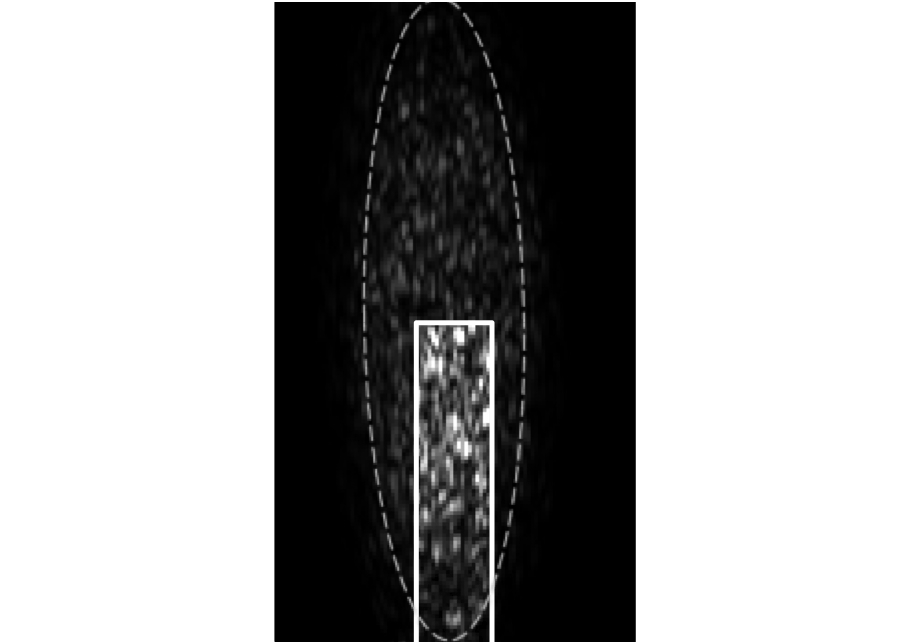}}
    \end{center}\caption{(a) Re-imaged at the CCD is the fibre far-field as it occurs at the slit plane at the spectrograph entrance. Shown are the far-fields (slit removed) at different strengths of fibre-agitation (from left to right): No agitation; agitation 1.5\,Hz; agitation 80\,Hz; additional manual agitation. Exposure time was 3\,s; (b) The fibre (far field) image can be truncated by a slit only as in fig.~\ref{fig:speckle_aperture} or by additional apertures that truncate the beam in cross-dispersion direction, e.g. when restricting the slit-aperture accordingly (as in section~\ref{sec:speck_stats}) or when overfilling the dispersive grating.}\label{fig:output}
\end{figure*}

\textit{Fibre input} (fig. \ref{fig:fiber_input}); The test fibre is illuminated using a power stabilized halogen lamp behind a  diffuser plate and a 500\,\textmu{}m pinhole. The lens system demagnifies the spot to a image size of 80 to 90\,\textmu{}m at the fibre input plane. An iris is used to stop the beam down in order to control the input focal ratio $F$. Focal ratios $F$ = 2.6, 3.9, 6.5 and 13.2 were used, by default $F=3.9$ if not stated otherwise. The high focal ratio of 3.9 is to mimic astronomical devices where the fibre is fed with a fast beam to minimize focal ratio degradation (FRD, \cite{Ramsey_1988}), an effect causing the output focal ratio $F_{out}$ to be generally smaller than the input focal ratio. Note that the focal ratio notation (F or f/\#) is applied here (and widely in optics literature) whereas often the geometrically more descriptive $\theta = $ arctan($2F$) is used in order to present the functional dependencies more clearly.\\
It is important here that the fibre core is fully illuminated. Choosing a significantly smaller spot-size results in poor scrambling. Incomplete scrambling manifests itself as a global structure of the intensity pattern in the spectrum (see \cite{Avila_2008}). The pattern is dependent on the position of the input spot, the variation exclusively occurs in cross-dispersion direction and thus differs from modal noise. This inhomogeneous intensity distribution is dependent of the position of the illuminating input spot. We chose a larger spot-size ($\approx$~70\,\% of the fibre-size) which resulted in a more homogeneous intensity pattern. A microscope is used to view the lateral position and the focus position of the pinhole image on the fibre core.\\
The beamsplitter is specified as "broadband non-polarizing" by manufacturer's specifications. Examining the beam at the fibre launch, it was shown that the beamsplitter is non-polarizing for the wavelength range under investigation. Utilizing a polarizing sheet before the beamsplitter, a polarization signal could not be detected beyond the detector noise, the beamsplitter was thus found to be non-polarizing (less than 10\% polarization for unpolarized input light).\\

\textit{Fibre agitation and slit unit} (fig. \ref{fig:fibre_agitation} and fig. \ref{fig:spec}); After the fibre input, the fibre is attached to a loudspeaker allowing the fibre to be agitated (for flat-field image acquisition). The agitation procedure can be understood as applying temporarily varying stress that is applied to the fibre. The refractive index in the fibre thus changes accordingly, causing the speckle pattern at the output to vary. For integration times $\gg$ the period of agitation the recorded image is a pattern that (ideally) does no longer contain any modal information. These are used for flat-fielding (see section \ref{sec:data_analysis}).\\
In order to obtain the most efficient agitation, the fibre is secured on two points where the second point of securing is chosen at a larger distance from the point of agitation (see fig.~\ref{fig:input_ag}). This allows the fibre to oscillate around its point of rest at this end. As it is believed that a large deflection changes the fibre properties by a larger amount, the distance is adjusted to the frequency at which it produces standing waves. An agitation frequency of ~80\,Hz (fig. \ref{fig:diff_agschemes}) gives the best agitation results (low $S\!N\!R$ in the respective exposures), and the modal pattern achieves a low contrast. Operating the agitation device at higher frequencies does not visibly improve the contrast, nor could a significant change in $S\!N\!R$ be observed. From laser exposures it appears that additional manual agitation would improve the result, but again no difference for the $S\!N\!R$-value could be measured.\\
The fibre end-face is placed behind a lens with focal length $f_0$ = 4\,mm, imaging the fibre far-field on the slit plane where the slit holder can carry slits of the widths 50, 100, 200 or 500\,\textmu{}m, realizing the spectral resolving powers as documented in Table\,\ref{tab:slit_width}. Note that placing a slit at that plane causes beam vignetting so that $\rho^2<1$, thus inducing modal noise (cf. eq. \ref{eq:GR}).\\

\begin{table} 
\caption{Slit width $s$, Resolving power $R$, and pixel per resolution element $p$  for the two cases of different focal length.}  
\centering      
\begin{tabular}{| c | c c | c c |}
\hline               
$s$ [\textmu{}m] & $R$ & $p$& $R$ & $p$\\ [0.5ex] 
\hline
50   & 228\,000 &2.3& 86\,000 &2.7\\   
100 & 114\,000 &4.6& 43\,000 &5.5\\ 
200 & 57\,000 &9.2& 21\,000 &11.0\\ 
500 & 23\,000 &23.0& 10\,000 &27.4\\ [1ex]	
	& \multicolumn{2}{|c|}{$f_1$ = \textit{800}$\,mm$}& \multicolumn{2}{|c|}{$f_1$ = \textit{300}$\,mm$}\\ [1ex]	
\hline     
\end{tabular} 
\label{tab:slit_width}  
\end{table} 

\textit{Echelle spectrograph} (fig. \ref{fig:spec}); The spectrograph features a 4f-configuration, popular in astronomical applications (image and pupil planes are situated in the focal planes of the imaging lenses), this has the advantage of homogeneous illumination of the grating, minimizing aberrations and allowing for stray light filtering. Using a lens with focal length $f_1=800$\,mm ($f_1=300$\,mm for lower resolution), the fibre end-face is re-imaged onto the Echelle grating. The Echelle grating is used in Blaze configuration, with the reflection angle $\psi=18^\circ$. A second lens of focal length $f_2=800$\,mm ($f_2=400$\,mm  for lower resolution) images the slit-restricted fibre far field onto the CCD. The spectral dispersion has been measured using a Neon lamp and the width of one spectral resolution element was determined using laser illumination. With these measurements the spectral resolving power can be calculated (see table~\ref{tab:slit_width}). A fold mirror is applied in order to fit the spectrograph on the optical bench. Mirrors and gratings are known to have polarizing characteristics and thus potentially attenuating a portion of the modes. Therefore the polarization was measured by introducing a polarizing sheet after the slit plane (see fig. \ref{fig:spec}) and found to be lower than 4\,\%. Together with the beam splitter, we concluded that the influence of polarizing effects resulting from mirror and grating are marginal and only lead to a small deviation from the non-polarized case (appendix B). Therefore, the $S\!N\!R$ in the following sections are the directly measured values, no polarization correction has been applied. Note that this needs to be taken into consideration when predicting modal noise effects in highly polarization filtering applications, e.g. when describing modal noise in polarimeters.\\
In order to capture the spectra, an Atik\,Instruments 314L+ camera was used. A narrow band filter (HeNe, $632.8\pm3.0$\,nm) is placed directly in front of the CCD to make sure that spectra of adjacent orders would not overlap. In addition, the slit-unit and spectrograph are placed in a light-tight box to keep the background signal low and air currents to a minimum.\\

\subsection{Data analysis}\label{sec:data_analysis}

\textit{1. Modal noise determination}\\
\begin{figure} 
  \centering
  \includegraphics[width=.5\textwidth]{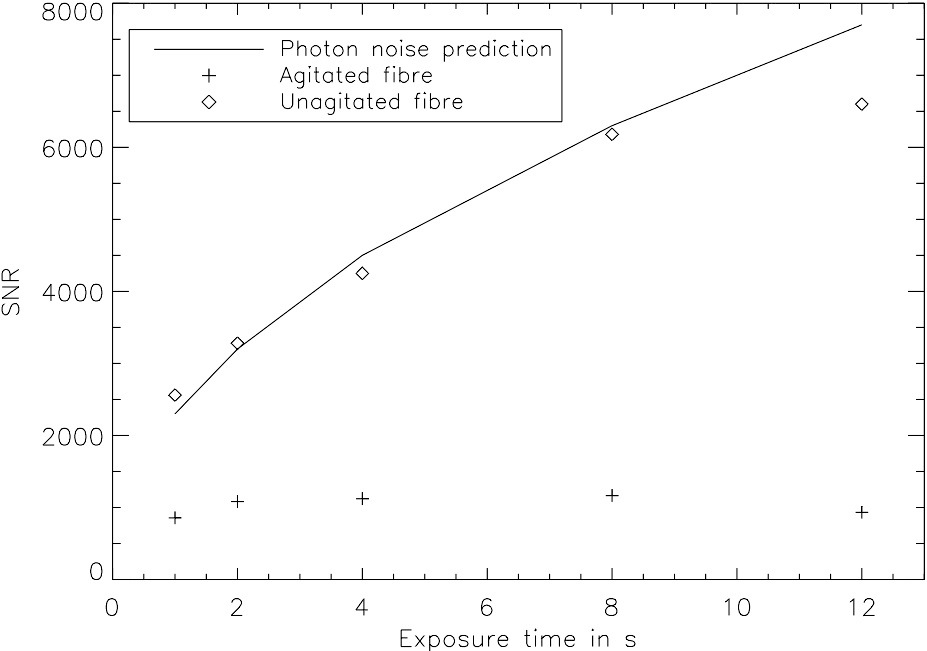}
    \caption{Exposure time was increased and for each dataset the modal noise calculated in order to find the point where the limiting signal-to-noise is reached. In this example 6\,s was chosen for the exposure time of the image stack, exposure times increase for higher signal-to-noise ratios.}
  \label{fig:SNR_sat}
\end{figure}

Each dataset consists of a set of exposures recording the fibre at rest (typically six images) containing the inherent modal noise signature of the bare fibre. These are supplemented by a set of exposures recording the fibre in the agitated state. Additionally, background images of the same exposure time are recorded. For short exposure times the $S\!N\!R$ reaches only low values, while longer exposure times result in increasing $S\!N\!R$ approaching the theoretical limiting $S\!N\!R$-level. Fig.~\ref{fig:SNR_sat} depicts how the exposure time was chosen: For increasing exposure times the data was recorded and its $S\!N\!R$ determined until a plateau value was reached. This data point is then used for further analysis. Choosing even longer exposure times diminishes the $S\!N\!R$ which indicates the occurrence of an additional source of noise. Very long exposure times can thus be critical and lead to a diminished $S\!N\!R$ that is overestimating the modal noise. In fact, we assume here that the limiting $S\!N\!R$-value is equal to modal noise, although other sources of noise can contribute significantly, especially in the high $S\!N\!R$-regime. Potential sources of noise are Poisson noise (eliminated for long exposures), variations in the light source intensity and colour as well as fluctuations in the speckle pattern during collection of the unagitated fibre spectra, which could be caused by air currents (minimized) or shift in temperature. The error in the $S\!N\!R$ of the unagitated fibre images of the measurement in fig.~\ref{fig:SNR_sat} is estimated during the data analysis process and of the order of 30-40, too small to appear in fig.~\ref{fig:SNR_sat}. In general,  $S\!N\!R$  measurements are conducted at least two or three times and their variation is taken as a measure for the uncertainty of the measurement and analysis process.\\

The data analysis involves basic reduction steps:
\begin{itemize}
\item{Background subtraction}
\item{Spectra extraction}
\item{Flat-fielding}
\item{Calculate standard deviation}
\end{itemize}

The background was subtracted using the corresponding dark-images. For exposure times longer than 10\,s the images exhibit hot pixels $\geq$ 3 times the average background signal; typically around five in the entire spectrum, despite the camera being cooled to 0$^\circ$C. These pixels are retrieved in the dark-images and removed in all exposures by replacing each with the mean-value of adjacent pixels.\\
We extracted the spectra by integrating over a large enough area in the cross-dispersion direction (300\,pixels) for focal ratios up to 200\,pixels full-width-half-maximum; the influence of background compared to the bright fibre-spectra is negligible. This could be shown from modal noise calculations using different sizes of integration ranges. Also, the read-out noise is $<$\,4\,ADU (i.e.\,\,1\,photo electron) per pixel (manufacturer's specifications) and is therefore neglected.\\
After spectra extraction we are left with a one-dimensional array containing intensity information over spectral position. In order to remove any instrumental signature (e.g. beating pattern due to the bandpass-filter), the unagitated fibre exposures are flat-fielded against the exposures of the agitated fibre. The quality of the flat-fielding is monitored by determining the modal noise for an independent set of images of the agitated fibre; This is compared with the actual modal noise as recorded by the unagitated fibre exposure.\\
The flat-fielded spectra are binned according to the size of one spectral resolution element (see Table \ref{tab:slit_width}). In the final step the standard deviation between the bins is calculated in order to obtain the signal to noise value. Note that in the case where the number of modes is small (as presented by \cite{Chen_2006}), the likelihood of measuring a certain intensity does not resemble the usual error function. In this case a Bayesian approach is necessary. However, as we are concerned here with the highly multi-mode case, the central limit theorem can be applied and the intensity values scatter in the familiar Gaussian distribution around the central value; i.e. the standard deviation is a valid approximation for the one sigma confidence level and error propagation law can be applied as usual.\\
After calculating the modal noise for the unagitated fibre exposures, the same procedure was applied to the exposures of the agitated fibre to compare with the theoretical photon noise, which was directly derived from Poisson statistics. With a few exceptions, all agitated fibre modal-noise values match with the theoretical photon noise, while always achieving significantly higher $S\!N\!R$-values than for the unagitated fibre. The remaining $S\!N\!R$-difference to the Poisson value is probably due to additional sources of noise (as discussed above).\\

\textit{2. Visibility determination}\\
\begin{figure} 
  \centering
  \includegraphics[width=.5\textwidth]{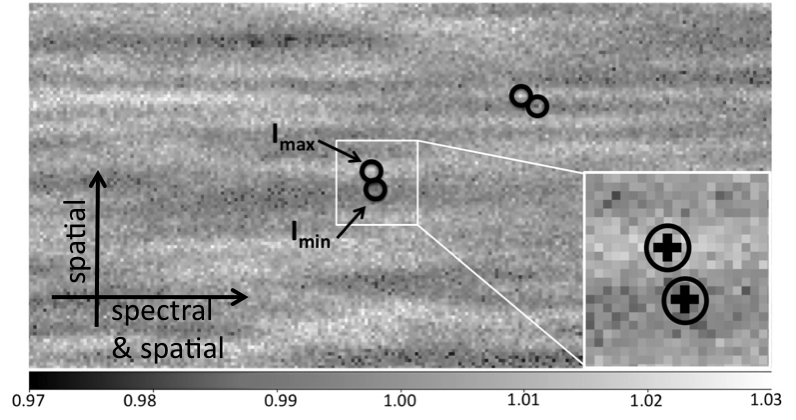}
    \caption{Unagitated fibre images are flat-fielded yielding a modal pattern. For visibility determination, adjacent minima and maxima are selected (circles) and the intensity of five pixel added (cross-shaped area) to feed into eq.\,(\ref{eq:vis}). The linear greyscale is normalized to the mean intensity.}
  \label{fig:visibility_det}
\end{figure}

The image in fig. \ref{fig:visibility_det} has been created by flat-fielding the central core area of the fibre-spectrum before applying the spectral extraction to the data. The pattern clearly shows a modal structure, the origin for modal noise. The visibility in each of these speckle patterns is derived by calculating the contrast between neighboring minima and maxima at six different positions of the spectrum: 
\begin{equation}\label{eq:vis}
v_{exp} = \frac{I_{max}-I_{min}}{I_{max}+I_{min}}.
\end{equation}
From these six visibility values the mean value is calculated, their scattering is used to estimate the uncertainty.


\section{Experimental results}\label{sec:exp_results}
\begin{figure} 
  \centering
  \includegraphics[width=.5\textwidth]{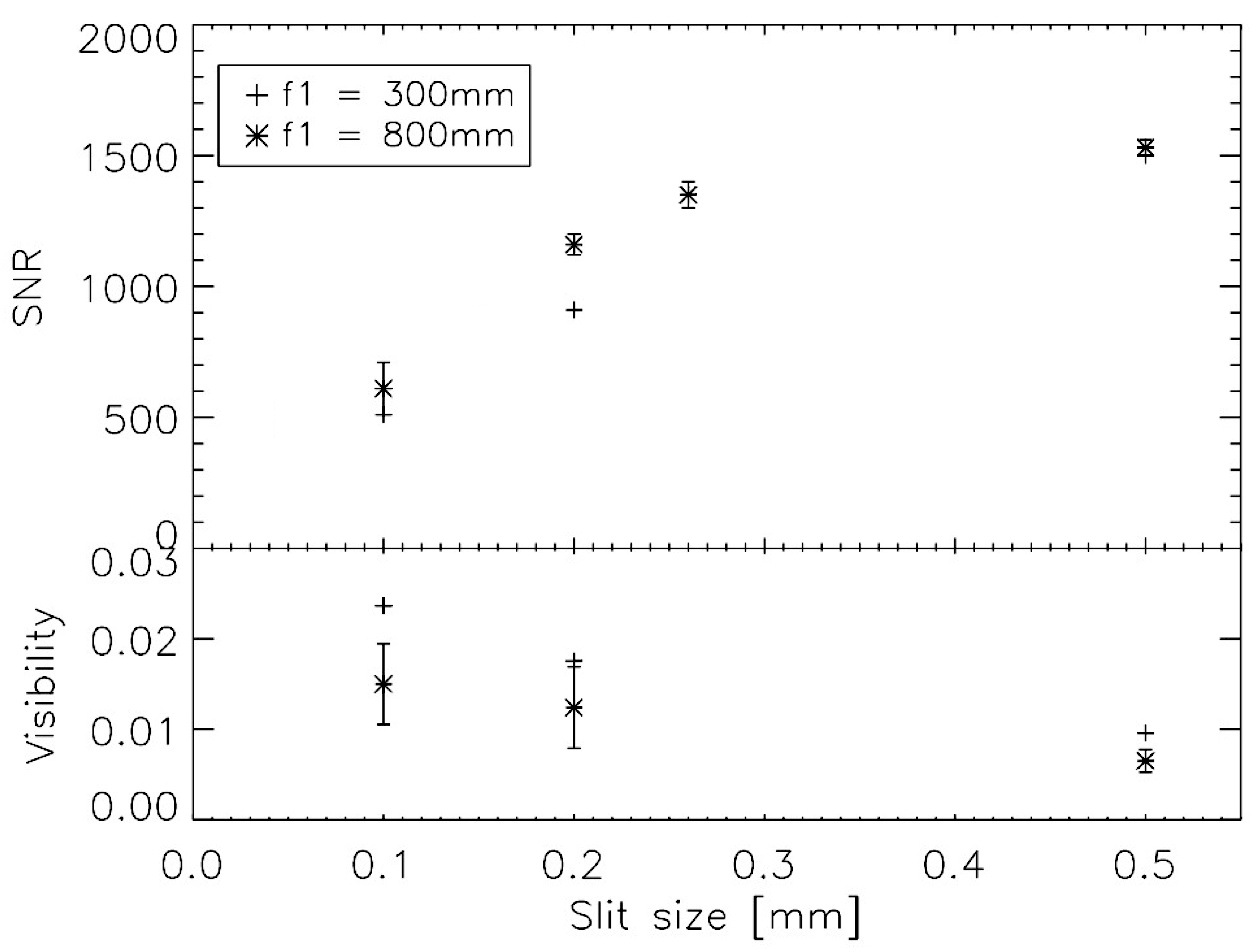}
    \caption{Signal to noise and visibilitity vs slit sizes $s$.}
  \label{fig:SNRraw_rhosq}
\end{figure}

\begin{figure} 
  \centering
  \includegraphics[width=.5\textwidth]{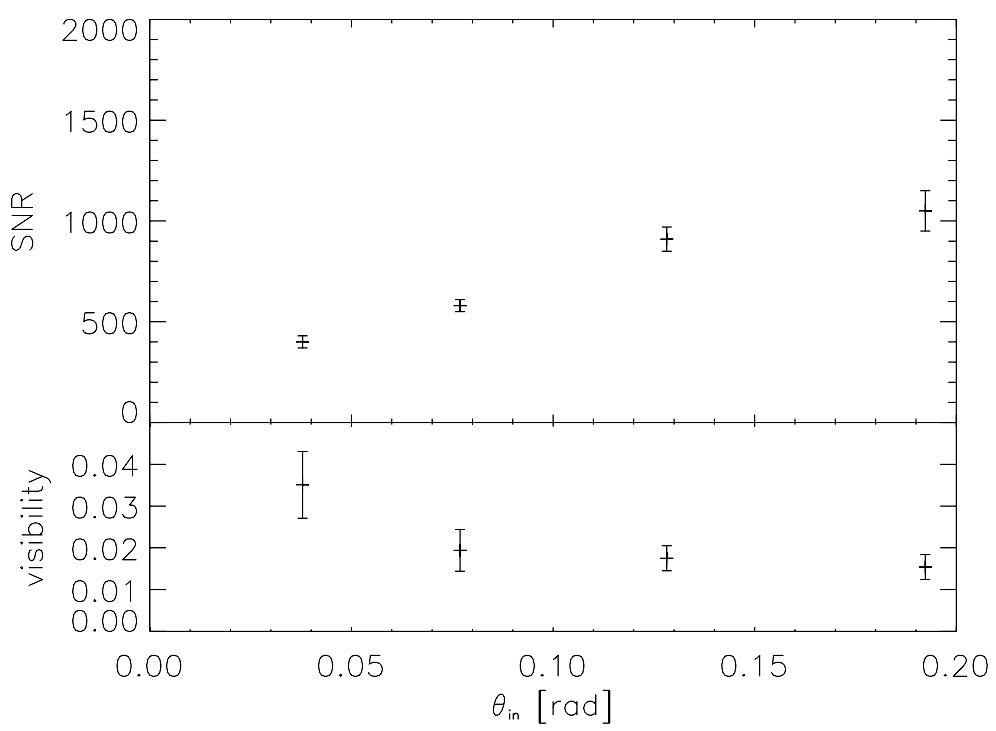}
    \caption{Signal to noise and visibility vs input focal ratio $\theta_{in}$.}
  \label{fig:SNRraw_thetain}
\end{figure}
Fig.~\ref{fig:SNRraw_rhosq} and fig.~\ref{fig:SNRraw_thetain} show slit size $s$ and input focal ratio $\theta_{in}$ vs.  signal-to-noise $S\!N\!R$ and visibility $v$, respectively. These values are directly derived as described in section \ref{sec:data_analysis}. For this particular setup, the $S\!N\!R$-values are of the order of $10^3$ and lower, and the visibility-values almost without exception do not exceed 5\%, typical values are found to be within 1-2\%. This is is significantly lower than any other data that can be found in the literature (e.g. \cite{Kanada_1983}).\\
As can be seen, variation of an input parameter (e.g. $s$ or $\theta_{in}$) changes both, $S\!N\!R$ and $v$. Eq.~(\ref{eq:GR_KA}) suggests that $S\!N\!R$ is composed of visibility and a coherent term, therefore determining the direct $S\!N\!R$-dependence is complicated. By separate inspection of these two terms we can measure $v$, but $S\!N\!R_{c\!o\!h}$ is difficult to measure and not directly ascertainable from the experimental data. However, rather than measuring each of its components separately, we divide the task of determining a concise $S\!N\!R$-model into two steps:
\begin{enumerate}
\item{\textit{Verifying the coherent model}: We start out with the assumption that  $S\!N\!R$  is factorisable. In order to show that this assumption is true, it is sufficient to show that $v\times S\!N\!R$ gives the same response as if illuminating the spectrograph with monochromatic light (from eq.~\ref{eq:GR_KA}: $S\!N\!R_{coh}$ = $v\times S\!N\!R$, this is the factorisation).}\label{item:coh_mod}
\item{\textit{Determine visibility dependence}: After showing that~\ref{item:coh_mod} is fulfilled, the problem of modal noise prediction reduces to the problem of predicting the visibility. The visibility needs to be investigated.}
\end{enumerate}

\subsection{Verifying the coherent model}\label{sec:speck_stats}
We show that the experimental results support that  $S\!N\!R$ is factorisable as stated in \ref{item:coh_mod} - $v\times S\!N\!R$ can be identified with $S\!N\!R_{coh}$ because the product follows the the functional dependence as described by eq.~(\ref{eq:GR}) for (a) $\rho^2$  and (b) the mode number $M$.
\begin{figure} 
  \centering
  \includegraphics[width=.5\textwidth]{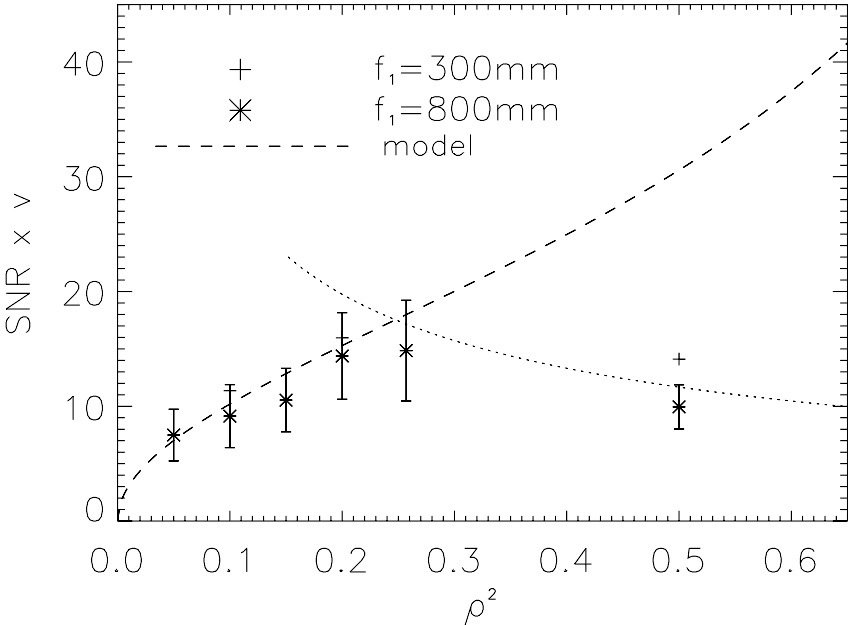}
    \caption{Signal-to-noise times visibility vs. $\rho^2$ for $f_1 =$ 300\,mm and 800\,mm data (error bars are in the same range for both data sets, but for clarity reasons only represented for the $f_1 =$ 800\,mm-case). The theoretical prediction corresponds to $M$ = 900 modes as can be obtained from an independent speckle-count measurement (see fig. \ref{fig:Meff300_thetain}). This is consistent with the six data points at $\rho^2 \leq $0.2. Deviations occur for  $S\!N\!R>1\,200$ (dotted line) as is the case for $\rho^2$ = 0.25 and 0.5, discussion see text.}
  \label{fig:SNRvis_rhosq}
\end{figure}

Fig.~\ref{fig:SNRvis_rhosq} shows ($v\times S\!N\!R$) vs $\rho^2$ ($\sim$ aperture area). The coherent model is supported by the six data points at $\rho^2 \leq 0.2$ of the high resolution ($f_1 =$ 800\,mm) and the low resolution ($f_1 =$ 300\,mm) setup. The theoretical prediction assumes a mode volume $M$ of 900\,modes. Not included into the theoretical fit is the $\rho^2 =$\,0.25 and 0.5 data points as their low value indicates a systematic deviation from the theory. We therefore expect an additional source of noise that becomes effective when approaching high  $S\!N\!R$ -values (of the order of 1\,200 and higher, see dotted line in fig.~\ref{fig:SNRvis_rhosq}) and longer exposure times (see also fig.~\ref{fig:SNR_sat} where the $S\!N\!R$ decreases after reaching a maximum) - For an investigation with lower $S\!N\!R$-values, higher $\rho^2$-values, a higher resolution setup or a test-fibre guiding fewer modes would be required. The error bars in fig.~\ref{fig:SNRvis_rhosq} are dominated by the uncertainty in visibility prediction.\\
The $\rho^2$-values are derived from the different slit sizes restricting the beam at the spectrograph entrance (compare fig. \ref{fig:speckle_aperture}):
\begin{equation}\label{eq:rhosq_F}
\rho^2 = \frac{A_d}{A_f} \approx \frac{sD}{(\pi/4) D^2} = \frac{4s}{\pi D} = \frac{4s}{\pi f_0/F_{out}} 
\end{equation}
where in the second step the approximation $A_d\approx sD$ has been used. With slit width $s$, diameter of the far field image $D$,  and $f_0$\,=\,4\,mm (focal length of the lens imaging the fibre far-field onto the spectrograph entrance slit). The input focal ratio is $F_{in}=3.9$. Due to FRD the corresponding output focal ratio  $F_{out}$ is 3.5. With eq. (\ref{eq:rhosq_F}) the available slit-widths 0.1, 0.2, 0.25 and 0.5\,mm thus correspond to $\rho^2$ = 0.1, 0.2, 0.25 and 0.5, respectively ($\rho^2 = $0.05 and 0.15 are realized by restricting the 0.1\,mm and the 0.2\,mm slit in cross-dispersion direction, see fig.~\ref{fig:other_rhosq}). The other input/output focal ratios are treated similarly.\\
In further investigation for the additional source of noise, we measured the $S\!N\!R$ for the no-slit case, this could not fully explain the additional noise that causes the $\rho^2=0.5$ data points to deviate from the predicted value. This in turn might indicate that this instability arises from the speckle pattern itself possibly being subject to changes for long integration times. A slight decline could be detected for longer integration periods and using a neutral density filter (20\,\% transmission) accordingly. Especially in the very sensitive high $S\!N\!R$-regime the impact can become very significant.

\begin{figure} 
  \centering
  \includegraphics[width=.5\textwidth]{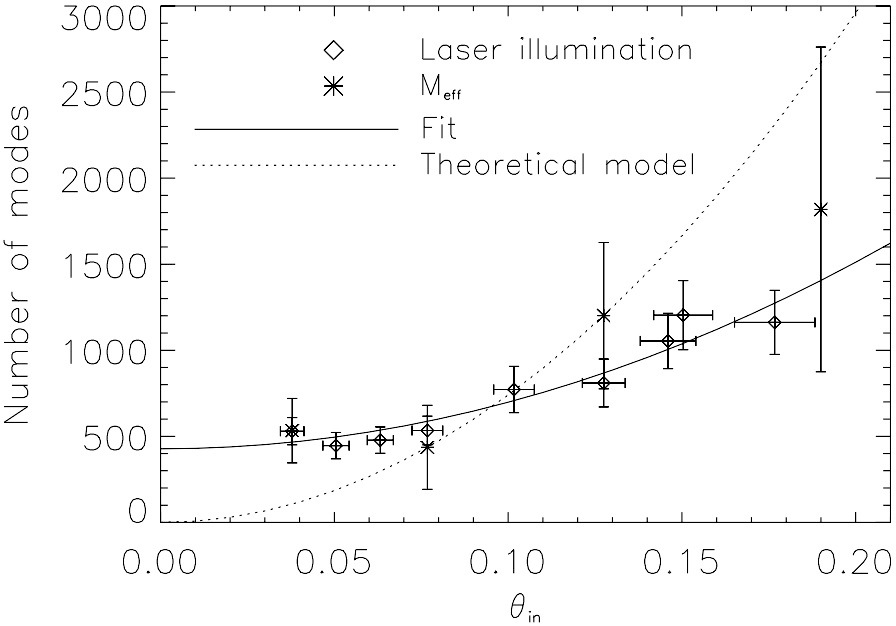}
    \caption{Number of guided modes vs.  $\theta_{in}$. $M_{e\!f\!f}$ is derived from modal noise and visibility measurements as in eq.~\ref{eq:mode_eff}. The dotted line represents the theoretical fit to $\propto\theta^2$ of the two central values at $\theta_{in} = $0.08 and 0.13. Counting speckles under laser illumination, the $\theta_{in}$-dependence obeys the square law (solid line) but with an offset.}
  \label{fig:Meff300_thetain}
\end{figure}	

To further approve the assumption in \ref{item:coh_mod}, i.e. $S\!N\!R_{c\!o\!h}$\,=\,$v\times S\!N\!R$ we investigate its $\theta$-dependence.  If $S\!N\!R$ is factorisable as proposed in eq.~(\ref{eq:GR_KA}) and if we believe in the $\rho^2$-dependence as stated; using eq.~(\ref{eq:GR}) the value $S\!N\!R\times v$ is a measure for the number of modes that are excited in the fibre at the central wavelength:
\begin{equation}
\label{eq:mode_eff}
M_{e\!f\!f}= (S\!N\!R\cdot v)^2\left(\frac{1-\rho^2}{\rho^2}\right)-1.
\end{equation}
This is a function of $\rho^2$ (known value, set by experiment) and the $S\!N\!R\times v$-product. Hence it follows $M_{e\!f\!f}$ = $M_{e\!f\!f}(S\!N\!R\times v)$.\\
Fig.~\ref{fig:Meff300_thetain} shows effective number of excited modes $M_{e\!f\!f}$ vs. angle of the input cone $\theta_{in}$, where the crosses are the $M_{e\!f\!f}$-values (see eq.~(\ref{eq:mode_eff})) which are derived from $S\!N\!R\times v$-values. The dashed line is a fit on the experimental data $\propto\theta_{in}^2$ and the diamonds represent the number of modes derived from laser illumination. The fit on the speckle data (solid line) obeys the square law, but has an offset. This implies that there will always be a certain number of modes, irrespective how small we choose the input angle $\theta_{in}$.\\
Note here that $\rho^2$ varies with the size of the far field at the output of the fibre, which is subject to input focal ratio and focal ratio degradation. Moreover, the term \textit{effective} mode number is used here to comply with the fact that the modes are not equally excited. The total number of excited modes is lower than the number of modes that is allowed for the mode volume in eq. (\ref{eq:mode_volume}). Depending on the symmetry of the fibre launch, different mode groups are excited.\\
According to \cite{Daino_1979}, the number of excited modes is equal to the number of speckles that can be counted when observing the fibre under monochromatic light. A standard He-Ne laser instead of the white light source. Here it has to be considered that the laser light is polarized, thus the expected number of speckles is only half compared to the case of unpolarized input light. The number of modes depicted in fig.~\ref{fig:Meff300_thetain} is therefore derived by multiplying the number of counted speckles by two.\\
The relationship between input focal ratios f/3.9 and f/6.5 ($\theta_{in}$ = 0.13 and 0.08) is conform with the theory, i.e. the number of  excited modes rises with the square of the coupling angle. 
However, the $\theta_{in}$ = 0.04 and 0.19 are inconsistent with the theory. However, all $M_{e\!f\!f}$-values show consistence with the number of modes as derived from counting the speckles under laser illumination, indicating that the product $S\!N\!R\times v$ can indeed be identified with the number of guided modes and thus supporting the factorization described in \ref{item:coh_mod}.\\
It is not straightforward to decide whether to base the model prediction on input focal ratio, output focal ratio or a mixture of both, because part of the focal ratio degradation could occur at the each of the fibre endfaces (compare length-effect, \cite{Poppett}), leading to a complex mode-population dependence along the fibre. We chose modal noise to be a function of $\theta_{in}$ as this seems to fit the coherent model best. For a better grounded statement, the actual mode population in the fibre as a function of the coupling angle needs to be known; Unfortunately, this is hard to determine.\\

\subsection{Determining visibility dependence}\label{sec:visibility}

\begin{figure} 
  \centering
  \includegraphics[width=.5\textwidth]{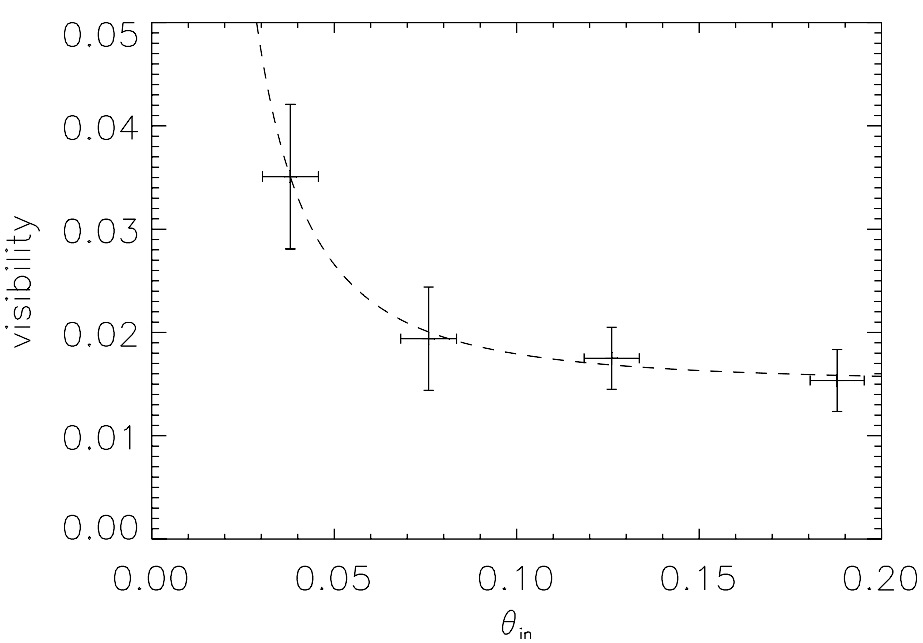}
    \caption{Visibility values for different input focal ratios. The experimental values well reflect a $\theta^{-2}$-law (dashed line).}
  \label{fig:vis300_inputfr}
\end{figure}

\begin{figure} 
  \centering
  \includegraphics[width=.5\textwidth]{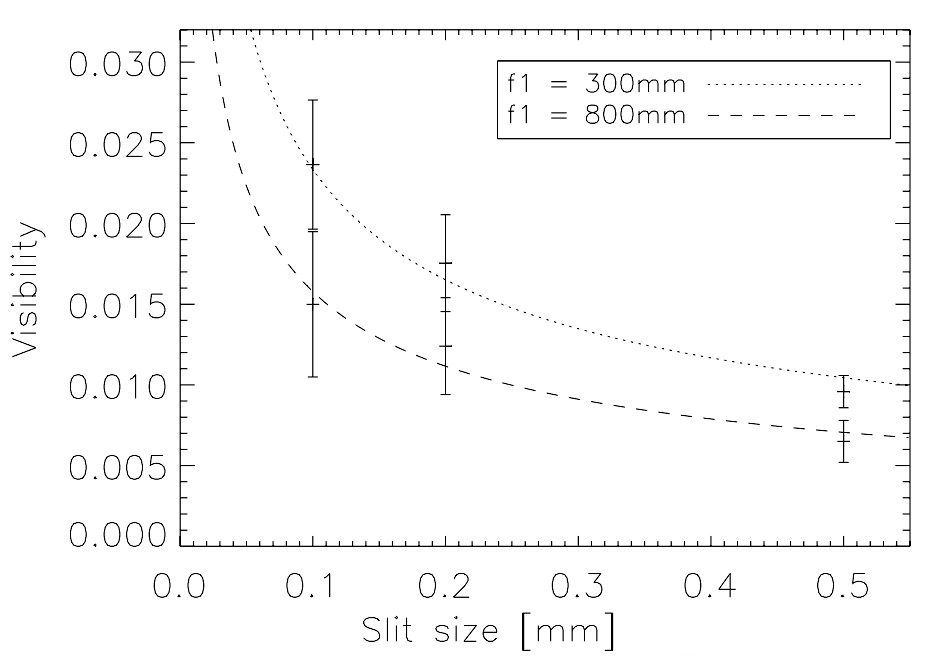}
    \caption{Visibility values for different slit widths, compared for two different resolving powers ($f_1$ = 300\,mm and 800\,mm).}
  \label{fig:vis300and800_slitsize}
\end{figure}

The goal of this section is to determine the visibility dependence of the spectrograph parameters; input focal ratio and slit-width at the spectrograph entrance. Previous considerations have shown that the initial hypothesis holds true; the visibility signal-to-noise product is consistent with the coherent model as described in eq. (\ref{eq:GR}). Furthermore, the number of excited modes rises with $\theta^2$ (fig.~\ref{fig:Meff300_thetain}) as determined from directly counting the number of speckles at the fibre exit (due to the offset slightly different from the theoretical prediction). Therefore we can conclude that, once the visibility is known, the modal noise is determined (see section \ref{sec:speck_stats}). Thus, the visibility is now subject to closer investigation. A fibre that guides a large number of modes will consequently possess a large range of propagation constants and thus will exhibit a low visibility. Likewise, a wide slit reduces visibility (cf. eq.~(\ref{eq:vis_est})).\\
The graph in fig.~\ref{fig:vis300_inputfr} shows the visibility as a function of the input focal ratio $\theta_{in}$ (same data as used in fig.\,\ref{fig:SNRraw_thetain} and fig.\,\ref{fig:Meff300_thetain}). From eq. (\ref{eq:vis_est}) we expect  the visibility to follow a  $\theta^{-2}$-law, which indeed reflects the experimental data if we accept an offset of 1.7\%. This offset becomes dominant for fast input beams ($\theta > 0.05$\,rad). Visibility as function of slit width is depicted in fig.~\ref{fig:vis300and800_slitsize}. Although eq.~(\ref{eq:vis_est}) predicts an $s^{-1}$-dependence, the experimental values suggest $s^{-0.5}$-dependence. These first results give us a parameterization, which reflects notably well the predictions of the simplified model. However, it also shows that the analytical model needs further refinement and may best be understood by simulation based investigations that reflect the complex physical processes when speckle images of adjacent wavelengths interfere at the detector plane.

Counter-intuitively, the supposed higher spectrograph resolving power for $f_1$ = 800\,mm (= higher coherence) shows lower visibility values than the lower resolving power setup ($f_1$ = 300\,mm). It cannot be excluded that this effect is an inherent feature of the way the spectra overlap at the detector. However, the authors hold the view that the apparent independence is more likely due to problems in the setup: repeated measurements with re-alignment of the fibre-launch resulted in a $S\!N\!R$-degradation of about 15\%; also differences in the optical performance of the applied collimator lenses possibly affect the measurement. In order to conclude for the visibility dependence on spectral resolving power this point needs clarification.\\

\section{Conclusions and outlook}\label{sec:conclusions}
Modal noise is a phenomenon of increasing importance as it poses a fundamental constraint on the achievable signal-to-noise of future high resolution spectroscopy applications. In this paper, special focus has been drawn on the characteristics of fibre-spectrographs used in astronomy, which we found to be insufficiently addressed in the literature. Furthermore, we experimentally verify that the model for modal noise prediction can be separated out into a visibility and a coherent term. Thus, modal noise prediction reduces mainly to the prediction of the visibility.\\
We have outlined the background understanding of modal noise as presented in existing literature, along with a simple ray-propagation model for visibility prediction. A test-bench for experimental investigations was built, 
with resolving powers in the range of $R=10\,000-200\,000$. Its design and data-analysis are presented. The results support the separation approach, showing that modal noise factorizes as predicted. As the coherent term can be determined from speckle-statistic considerations, it is sufficient to know the visibility for noise prediction. Therefore, our further examinations concentrated on predicting the visibility. The functional dependence of the experimentally derived visibility values show partial agreement with the simplified theoretical description. These considerations need yet to be extended by further simulation work, while carefully paying attention to the spectrograph geometries. This rather extensive simulation work will potentially complete the modal noise picture which will be examined in a future paper.

\appendix
\section{Visibility estimation}

The bandwidth $f_c$ in eq.~\ref{eq:KA} can be estimated as follows: Assuming a light cone of focal ratio $F = 1/2\theta$ is coupled into the fibre, the difference in propagation constant between on-axis rays ($\beta_0$) and rays coupled with the angle $\theta$ (propagation constant $\beta$) is given by
 \begin{eqnarray*}
\Delta\beta=\beta_0-\beta_0\sqrt{1-\theta^2/n_{c}^2} \approx \frac{1}{2}\frac{\theta^2}{n_{c}^2}\beta_0=\frac{1}{2}\frac{\theta^2}{n_{c}^2}\frac{2\pi\nu}{c/n_{c}},
 \end{eqnarray*}
 where $n_{c}$ is the core refractive index, c the speed of light in vacuum and $\nu$ is the frequency of the electromagnetic wave, $\beta_0=2\pi\nu/(c/n_{c})$ has been used.\\
 The time needed for a mode traveling the whole fibre length $L$ is given by $t=\beta L/2\pi\nu$. Accordingly, the time delay between on-axis and off-axis rays (launched at $\theta$) is then $\Delta t=\Delta\beta L/2\pi\nu$. Identifying this time delay with $f_c^{-1}$ yields
\begin{eqnarray*}
f_c = \frac{2\pi\nu}{\Delta\beta L}= \frac{2n_{c}c}{L\theta^2}.
\end{eqnarray*}
An estimation for $f_s$ is derived as follows;
\begin{eqnarray*}
f_s =\nu_1-\nu_2\approx\frac{\nu_c}{R} =  \frac{c}{R\lambda_c},
\end{eqnarray*}
Where $\nu_c$ and $\lambda_c$ are the central frequency and the central wavelength of the spectral resolution element with width $\Delta\nu=\nu_1-\nu_2$. 
Hence,
\begin{eqnarray*}
f_s/f_c = \frac{L\theta^2}{2n_{c}R\lambda}
\end{eqnarray*}
can be substituted in eq. (\ref{eq:KA}), yielding
\begin{eqnarray*}
v = \left[1+2\left(\frac{L\theta^2}{2n_{c}R\lambda}\right)^2\right]^{-1/2}. 
\end{eqnarray*}
Note: It has to be said that skew rays are omitted here for simplification.
\section{Modal noise for partially polarized light}

For a circular fibre the number of excited modes is equal in both polarization planes:
\begin{equation}\label{eq:Pol_SNR}
M_\parallel=M_\perp=\frac{M}{2}.
 \end{equation}
The two speckle patterns for each polarization plane interfere independently and are integrated additively at the detector. We can ascribe two independent signal-to-noise values to each of these patterns.
\begin{equation}
S\!N\!R_{\parallel,\perp} = a\cdot\sqrt{M_{\parallel,\perp}}
 \end{equation}
where $a = \rho/(v\sqrt{1-\rho^2})$. If the intensities of these two planes differ (partial linear polarization) by
 \begin{equation}
 I_\parallel - I_\perp= p\cdot I
  \end{equation}
(where $p$ is polarization degree); the combined $S\!N\!R$ results in
 \begin{eqnarray}
S\!N\!R	&=&\sqrt{S\!N\!R_\parallel^2+(1-p)S\!N\!R_\perp^2}  = \\\nonumber
		&= &a\sqrt{(M_\parallel+(1-p)M_\perp)} = a\sqrt{1-\frac{p}{2}}\sqrt{M}.
  \end{eqnarray}
  In the last step eq.~\ref{eq:Pol_SNR} has been used. For 10\,\% polarized light, the measured $S\!N\!R$ will be reduced by $1-\sqrt{1-0.1/2}\approx$ 2~\% compared to non-polarized light.
  
  \end{document}